\documentstyle [12pt] {article}

\newcommand{\beq}{\begin{equation}}
\newcommand{\eeq}{\end{equation}}
\newcommand{\beqs}{\begin{eqnarray}}
\newcommand{\eeqs}{\end{eqnarray}}

\begin{document}

\def\thefootnote{\fnsymbol{footnote}}
\baselineskip 6.0mm

\begin{flushright}
\begin{tabular}{l}
ITP-SB-95-11    \\
May, 1994
\end{tabular}
\end{flushright}

\vspace{5mm}
\begin{center}
{\Large \bf A Connection Between Complex-Temperature}

\vspace{2mm}

{\Large \bf Properties of the 1D and 2D Spin $s$ Ising Model}

\vspace{12mm}

\setcounter{footnote}{0}
Victor Matveev\footnote{email: vmatveev@max.physics.sunysb.edu}
\setcounter{footnote}{6}
and Robert Shrock\footnote{email: shrock@max.physics.sunysb.edu}

\vspace{6mm}
Institute for Theoretical Physics  \\
State University of New York       \\
Stony Brook, N. Y. 11794-3840  \\

\vspace{12mm}

{\bf Abstract}
\end{center}
    Although the physical properties of the 2D and 1D Ising models are
quite different, we point out an interesting connection between
their complex-temperature phase diagrams.  We carry out an exact
determination of the complex-temperature phase diagram for the 1D Ising model
for arbitrary spin $s$ and show that in the $u_s=e^{-K/s^2}$ plane
(i) it consists of $N_{c,1D}=4s^2$ infinite regions separated by an equal
number of boundary curves where the free energy is non-analytic;
(ii) these curves extend from the origin to complex
infinity, and in both limits are oriented along the angles $\theta_n =
(1+2n)\pi/(4s^2)$, for $n=0,..., 4s^2-1$; (iii) of these curves, there are
$N_{c,NE,1D}=N_{c,NW,1D}=[s^2]$ in the first and second (NE and NW)
quadrants; and (iv) there is a boundary curve (line) along the negative real
$u_s$ axis if and only if $s$ is half-integral.  We note a close relation
between these results and the number of arcs of zeros protruding into the FM
phase in our recent calculation of partition function zeros for the 2D spin $s$
Ising model.

\vspace{16mm}

\pagestyle{empty}
\newpage

\pagestyle{plain}
\pagenumbering{arabic}
\renewcommand{\thefootnote}{\arabic{footnote}}
\setcounter{footnote}{0}

    Recently, we presented calculations of the complex-temperature (CT)
zeros of the partition functions for square-lattice Ising models with several
higher values of spin, $s=1, \ 3/2$, and 2 \cite{hs}.  In the thermodynamic
limit, these zeros merge to form curves across which the free energy is
non-analytic, and thus calculations for reasonably large finite lattices
give insight into the complex-temperature phase diagrams of these models.
These phase diagrams consist of the complex-temperature extensions of the
Z$_2$--symmetric, paramagnetic (PM) phase; of the two phases in which the
Z$_2$ symmetry is spontaneously broken with long-range ferromagnetic (FM)
and antiferromagnetic (AFM) order; and, in addition, certain
phases which have no overlap with any physical phase (denoted ``O'' for other).
Some of the zeros lie along curves which, in the thermodynamic limit,
separate the various phases.  In addition, there are zeros lying along
various curves or arcs which terminate in the interiors of the FM and AFM
phase.  Physical and CT singularities of the magnetization,
susceptibility, and specific heat obtained from analysis of low-temperature
series have been discussed recently for the square lattice Ising model with
the higher spin values $s=1$ \cite{egj} and $s=1$, 3/2, 2, 5/2, and 3
\cite{jge}.

    Here we give some further insight into the complex-temperature phase
diagrams for higher-spin Ising models.  We first report an exact
determination of the CT phase diagrams of the 1D Ising model
for arbitrary spin $s$.  We then point out a very interesting connection
between features of these 1D phase diagrams and certain properties of
the phase diagrams of the Ising model on the square lattice inferred from our
calculation of partition function zeros for higher spin values.  This
connection is useful because, unlike the 2D spin 1/2 case, no exact
closed--form
solution has ever been found for the 2D Ising model with spin $s \ge 1$, and
hence further elucidation of its properties is of continuing value, especially
insofar as these constrain conjectures for such a solution.
Of course, the physical properties of a spin model at its lower critical
dimensionality (here $d_{\ell . c.d.}=1$) are quite different from
those for $d > d_{\ell . c. d.}$.  However, as we shall discuss, some of the
properties of the CT phase diagram for $d=2$ exhibit simple
relations with the $d=1$ case. \footnote{Indeed, from the $d=1+\epsilon$ and
$d=2+\epsilon$ expansions for the Ising and O($N$) models \cite{eps}, one
knows that expansions above $d_{\ell . c. d.}$ can even give useful
information about physical critical behavior.}

   There are several reasons why CT properties of statistical
mechanical models are of interest.  First, one can understand more deeply the
behavior of various thermodynamic quantities by seeing how they behave as
analytic functions of complex temperature.  Second, one can see how the
physical phases of a given model generalize to regions in appropriate
CT variables. Third, a knowledge of the CT
singularities of quantities which have not been calculated exactly helps in
the search for exact expressions for these quantities.  Fourth,
one can see how CT singularities in functions such as the
magnetization and susceptibility are associated with the boundaries of the
phases and with other points where the free energy is
non-analytic.  Such CT properties were first considered (for
the 2D, $s=1/2$ square-lattice Ising model) in Ref. \cite{fisher} and for
higher-spin (2D and 3D) Ising models in Ref. \cite{fg}.

    The spin $s$ (nearest-neighbor)
Ising model is defined, for temperature $T$ and external
magnetic field $H$, by the partition function
$Z = \sum_{\{S_n\}} e^{-\beta {\cal H}}$ where, in a commonly used
normalization,
\beq
{\cal H} = -(J/s^2) \sum_{<nn'>} S_n S_{n'} - (H/s) \sum_n S_n
\label{ham}
\eeq
where $S_n \in \{-s, -s+1,...,s-1,s\}$ and $\beta = (k_BT)^{-1}$. $H=0$
unless otherwise indicated. We define $K = \beta J$
and $u_s = e^{-K/s^2}$.  $Z$ is then a generalized (i.e. with
negative as well as positive powers) polynomial in $u_s$.
The (reduced) free energy is
$f = -\beta F = \lim_{N_s \to \infty} N_s^{-1} \ln Z$ in the thermodynamic
limit.

   For $d=1$, one can solve this model by transfer matrix methods. One has
\beq
Z = Tr({\cal T}^N) = \sum_{j=1}^{2s+1} \lambda_{s,j}^N
\label{zeq}
\eeq
where the $\lambda_{s,j}$, $j=1,...2s+1$ denote the eigenvalues of the
transfer matrix ${\cal T}$ defined by
${\cal T}_{nn'}=<n|\exp((K/s^2)S_n S_{n'})|n'>$
(we assume periodic boundary conditions for definiteness).  It is
convenient to analyze the phase diagram in the $u_s$ plane. For physical
temperature, phase transitions are associated with degeneracy of leading
eigenvalues \cite{al}.
There is an obvious generalization of this to the case of complex
temperature: in a given region of $u_s$, the eigenvalue of ${\cal T}$
which has maximal magnitude, $\lambda_{max}$,  gives the
dominant contribution to $Z$ and hence, in the thermodynamic
limit, $f$ receives a contribution only from $\lambda_{max}$:
$f=\ln ( \lambda_{max})$.  For complex $K$, $f$ is, in general,
also complex.  The CT phase boundaries are determined by the degeneracy, in
magnitude, of leading eigenvalues of ${\cal T}$.  As will be evident in our 1D
case, as one moves from a region with one dominant eigenvalue
$\lambda_{max}$ to a region in which a different eigenvalue $\lambda_{max}'$
dominates, there is a non-analyticity in $f$ as it switches from
$f=\ln (\lambda_{max})$ to $f=\ln (\lambda_{max}')$.  The boundaries of these
regions are defined by the degeneracy condition
$|\lambda_{max}|=|\lambda_{max}'|$.  These form curves in the $u_s$
plane.\footnote{By ``curves'' we include also the special case of a line
segment.}

   Of course, a 1D spin model with finite-range interactions has no
non-analyticities for any (finite) value of $K$, so that, in particular,
the 1D spin $s$ Ising model is analytic along the positive real $u_s$ axis.
For a bipartite lattice, $Z$ and $f$ are invariant under $K \to -K$, i.e.,
$u_s \to 1/u_s$.  It follows that the CT phase diagram also has this
symmetry, i.e., is invariant under inversion about the unit circle in the
$u_s$ plane.  This symmetry also holds for a finite bipartite lattice; for
$d=1$, the lattice is bipartite iff $N$ is even, and for our comments about
finite-lattice results, we thus make this restriction.  Further,
since the $\lambda_{s,j}$ are analytic functions of $u_s$, whence
$\lambda_{s,j}(u_s^*)=\lambda_{s,j}(u_s)^*$, it follows that the solutions to
the degeneracy equations defining the boundaries between different phases,
$|\lambda_{s,j}|=|\lambda_{s,\ell}|$, are invariant under $u_s \to u_s^*$.
Hence, the complex-temperature phase diagram is invariant under
$u_s \to u_s^*$.

   We shall present results for a few $s$ values explicitly.  For $s=1/2$,
one has $(u_{1/2})^{1/4}\lambda_{1/2,j}= 1 \pm (u_{1/2})^{1/2}$.  $f$ is an
analytic function of $u_{1/2}$ except at
points which constitute the solution to
$|\lambda_{1/2,1}|=|\lambda_{1/2,2}|$; these comprise the
negative real axis, $-\infty \le u_{1/2} \le 0$.  Apart from this line, the
dominant eigenvalue of ${\cal T}$ is $\lambda_{1/2,1}$.
For $s=1$, the eigenvalues of $T_s$ are $\lambda_{1,1}=u_1^{-1}-u_1$ and
\beq
\lambda_{1,j=2,3}=(1/2)\Bigl [ u_1^{-1}+1+u_1 \pm
(u_1^{-2}-2u_1^{-1}+11-2u_1+u_1^2)^{1/2} \Bigr ]
\label{lambda123}
\eeq
As shown in Fig. 1(a), the phase diagram consists of four phases, the
complex-temperature extension of the PM phase, together with three O phases.
The curves separating these phases are the solutions of
$|\lambda_{1,1}|=|\lambda_{1,2}|$.  The third eigenvalue, $\lambda_{1,3}$, is
always subdominant.  In the two phases containing the real $u_s$ axis,
$\lambda_{1,2}$ has maximal magnitude, while in the two containing the
imaginary $u_s$ axis, $\lambda_{1,1}$ is dominant.  The CT zeros of $Z$
calculated for finite lattices lie on or close to these curves, starting a
finite distance from the origin and being distributed in a manner symmetric
under the inversion $u_s \to 1/u_s$.  As the lattice size increases, the zeros
spread out, the one with smallest (largest) magnitude moving closer to (farther
from) the origin.  For $s=3/2$, the eigenvalues are given by
\beqs
\lefteqn{2u^{9/4}\lambda_{3/2,j} = (1+u^2)(1+\eta u^{5/2})} \nonumber \\
                        & & \zeta \Bigl [
(1-2u^2+4u^3+u^4+u^5+4u^6-2u^7+u^9)-2\eta u^{5/2}(1-6u^2+u^4) \Bigr ]^{1/2}
\label{lambda1p5j}
\eeqs
where here $u \equiv u_{3/2}$ and
$(\eta,\zeta)=(+,+), \ (-,+), \ (+,-), \ (-,-)$ for $j=1,2,3,4$.  The
phase diagram is shown in Fig. 1(b) and consists of nine regions separated by
the curves where $|\lambda_{3/2,1}|=|\lambda_{3/2,2}|$.  In the region
containing the positive real $u_s$ axis, $\lambda_{3/2,1}$ is dominant, and
as one makes a circle around the origin, each of the nine times that one
crosses a boundary, there is an alternation between $\lambda_{3/2,1}$ and
$\lambda_{3/2,2}$ as the dominant eigenvalue.

   We find the following results for general $s$: (i) the complex-temperature
phase diagram consists of
\beq
N_{c,1D}=4s^2
\label{nc1d}
\eeq
(infinite) regions separated by an equal number of boundary curves where the
free energy is non-analytic; (ii) these curves extend from the origin to
complex infinity, and in both limits are oriented along the angles
\beq
\theta_n = \frac{(1+2n)\pi}{4s^2}
\label{thetan}
\eeq
for $n=0,...4s^2-1$; (iii) of these curves, there are
\beq
N_{c,NE,1D}=N_{c,NW,1D}=[s^2]
\label{ncnenw}
\eeq
in the first and second (NE and NW) quadrants, where $[x]$ denotes the integral
part of $x$; and (iv) there is a boundary curve (which in this case is a
straight line) along the negative real $u_s$ axis if and only if $s$ is
half-integral.  ($N_{c,SE,1D}=N_{c,NE,1D}$ and
$N_{c,SW,1D}=N_{c,NW,1D}$ by the $u \to u^*$ symmetry.)
To derive these results, we carry out a Taylor series
expansion of the $\lambda_{s,j}$ in the vicinity of $u_s=0$.  The dominant
eigenvalues have the form
\beq
u_s^{s^2}\lambda_{s,j} = 1 + ... + a_{s,j}u_s^{2s^2} + ...
\label{lambdaexpansion}
\eeq
where the first $...$ dots denote terms which are independent of $j$ and the
second $...$ dots represent higher order terms which are dependent upon $j$.
Setting $u_s=r e^{i \theta}$ and solving
the equation $|\lambda_{s,j}|=|\lambda_{s,\ell}|$ yields $\cos(2s^2 \theta)=0$,
whence
\beq
2s^2\theta = \frac{\pi}{2} + n\pi \ , \qquad n=0,...4s^2-1
\label{thetasol}
\eeq
Each of these solutions
yields a curve across which $f$ is non-analytic, corresponding to a switching
of dominant eigenvalue.  These curves cannot terminate since if they did, one
could analytically continue from a region where the expression for $f$ depends
on a given dominant eigenvalue, to a region where it depends on a different
eigenvalue.  Owing to the $u_s \to 1/u_s$ symmetry of the model, a given
curve labelled by $n$ approaches complex infinity in the same direction
$\theta_n$ as it approaches the origin.  This yields (i) and (ii).  From
(\ref{thetan}), it follows that
\beq
0 < \theta_n < \pi/2 \ \qquad {\rm for} \ \ 0 \le n < [s^2-1/2]
\label{nequad}
\eeq
which comprises $[s^2]$ values, and similarly,
\beq
\pi/2 < \theta_n < \pi \ \qquad {\rm for} \ \ [s^2-1/2] < n < [2s^2-1/2]
\label{nwquad}
\eeq
which again
comprises $[s^2]$ values.  Finally, if and only if $s$ is half-integral, then
the equation $\theta_n=\pi$ has a solution (for integral $n$), viz.,
$n=(2s+1)(2s-1)/2$.  From the $u_s \to u_s^*$ symmetry of the phase diagram,
the curve corresponding to this solution must lie on the negative real axis for
all $r$, not just $r \to 0$ and $r \to \infty$. This yields (iii)-(iv).
\ $\Box$.

   We note that from (\ref{thetan}), the angular size of each
region near the origin (or infinity) is
\beq
\Delta \theta = \frac{\pi}{2s^2}
\label{deltatheta}
\eeq
Also, from the $u_s \to u_s^*$ symmetry of the phase diagram, it follows in
particular, that for each curve starting out from the origin at $\theta_n$,
there is a complex
conjugate curve at $-\theta_n$.  As is true of any theory at its
lower critical dimensionality, the model is singular at $K=\infty$, i.e,
$u_s=0$.   If and only if $s$ is integral, the $n$'th curve has another,
$m$'th curve which is related to it by $\theta_m
= \theta_n + \pi$ (whence $m=n+2s^2$), so that as one travels through the
origin on the $n$'th curve, one emerges on the other side on the $m$'th curve.

   As one can see from Fig. 1, as the boundary curves approach the unit circle
$|u_s|=1$, some of them twist in $s$-dependent ways.  Interestingly, several of
the points where they cross the unit circle coincide with points which we
inferred to be likely multiple (=intersection) points of the boundary curves of
the complex-temperature phase diagrams for the corresponding spin $s$ 2D Ising
model.  For example, for the 1D $s=1$ model, the $n=0$ and $n=1$ curves cross
the unit circle at $u_1=i$ and $u_1=e^{2 i\pi/3}$, respectively, and hence the
complex-conjugate curves ($n=3$ and $n=2$) cross this circle at $u_1=-i$ and
$u_1=e^{-2 i \pi/3}$.  The points $u_1 = e^{\pm 2 i\pi/3}$ are precisely the
values which we inferred for the intersection points of boundary curves in the
second (``northwest''=NW) and third (SW) quadrants of the 2D $s=1$ phase
diagram from our calculation of partition function zeros \cite{hs}.
For the 1D $s=3/2$ case, the $n=0,1,2$
and 3 curves cross the unit circle at the respective points
$u_{3/2}=e^{i \pi/3}$, $e^{2 i \pi/5}$, $i$, and $e^{4 i \pi/5}$, and so forth
for the complex conjugate curves.  Among these, the points $\pm i$ and
$e^{\pm 4 i \pi/5}$ are points inferred as likely intersection points of phase
boundaries in the NW and SW quadrants of the $u_{3/2}$ plane for the 2D
$s=3/2$ model \cite{hs}.  Similar correspondences hold for $s=2$.
These are intriguing results.

   The spin 1/2 Ising model is equivalent to the two-state Potts model.  While
determining the CT phase diagram of the 1D spin $s$ Ising model, it is thus
also of interest to compare this with that of the 1D $q$-state Potts model,
defined by $Z_P = \sum_{\sigma_n}e^{-\beta {\cal H}_P}$ with
${\cal H}_P = -J_P \sum_{<nn'>}\delta_{\sigma_n \sigma_{n'}}$
where $\sigma_n \in \{1,...,q\}$.  We define $K_P = \beta J_P$ and $u_P =
e^{-K_P}$. \footnote{Recall that the equivalence of ${\cal H}_P$ for $q=2$ to
(\ref{ham}) for $s=1/2$ entails $K_P=2 K$.}
The eigenvalues of the transfer matrix are $\lambda = u_P^{-1}-1$ ($q-1$
times) and $\lambda' = u_P^{-1} + (q-1)$.  Setting $u_P=re^{i\theta}$, the
equation for $|\lambda|=|\lambda'|$ is
\beq
q \Bigl ( (q-2)r + 2\cos \theta \Bigr ) = 0
\label{pottseq}
\eeq
For $q=2$, the solution is
the imaginary axis, $u_P=\pm i r$, equivalent to the negative real axis in
the $u_{1/2}$ variable.  For $q \ne 2$ (and $q \ne 0$ ), the solution is
$\cos \theta = -(q-2)r/2$ for $0 \le r \le 2/(q-2)$, i.e.,
\beq
u_P = \frac{-1+e^{i\omega}}{q-2}
\label{pottscircle}
\eeq
for $0 \le \omega < 2\pi$, viz., a circle centered at $u_P=-1/(q-2)$ with
radius $1/(q-2)$.  Thus, for all $q > 2$,
the complex-temperature phase diagram is qualitatively the same, consisting
of a PM phase containing the positive real $u$ axis, and an O phase inside
the circle (\ref{pottscircle}).\footnote{One may also formally consider
other values of $q$, as in the 2D Potts model.  In particular, for $q=0$, all
eigenvalues are identically equal, and the phase diagram is trivial.}
This is quite different from the 1D spin $s$ Ising phase diagram, for which
the number of phases does depend on $s$. This difference can
be traced to the fact that the structure of the transfer matrix is
qualitatively the same for different $q$ in the Potts model (all diagonal
entries equal to $u^{-1}$ and all off-diagonal entries equal to 1), whereas it
depends on $s$ in the Ising model.

   In addition to the correspondences already above, we have found a further
very interesting connection between the features of the 1D spin $s$ Ising
model and certain properties which we have observed from our calculations
of partition function zeros for the 2D (square-lattice) Ising model with
$s=1$, 3/2, and 2 \cite{hs}.
One of the interesting features which we found in the 2D case
was a certain number of finite arcs which protrude into the FM phase.  (By the
$u_s \to 1/u_s$ symmetry, there are also corresponding arcs protruding into the
AFM phase).  By combining our results with low-temperature series
analyses in Refs. \cite{egj,jge}, we observed that the divergences of the
spontaneous magnetization $M$ (which also imply divergences in the
susceptibility \cite{chitri}) occur at the endpoints of these arcs. The arc
endpoints are thus of considerable interest for the study of CT singularities.

   For the cases $s=1$, 3/2, and 2, the numbers of arcs
protruding into and terminating in the first and second (NE and NW) quadrants
of the (complex-temperature extension of the) FM phase are given in terms of
the numbers of boundary curves of the corresponding 1D spin $s$ model as
\beq
N_{e,NE,2D}=N_{c,NE,1D}-1=[s^2]-1
\label{narcsne}
\eeq
and
\beq
N_{e,NW,2D}=N_{c,NW,1D}=[s^2]
\label{narcsnw}
\eeq
so that the total number of arcs protruding into the FM (equivalently,
AFM) phase is $4[s^2]-2$.  Our derivations in the present paper thus provide
the explanation underlying our conjecture \cite{hs} that the total number of
arcs protruding into the FM (equivalently AFM) phase for the Ising model on the
square lattice with arbitrary spin $s$ is
\beq
N_{e,FM}=N_{e,AFM}=\max \{ 0,4[s^2]-2\}
\label{narcfm}
\eeq
where we have incorporated also the exact result that $N_{e,FM}=N_{e,AFM}=0$
for $s=1/2$.

   One can qualitatively describe how the complex-temperature phase
diagram changes as one goes from $d=1$ to $d=2$ as follows.
For $d > 1$, the
model is analytic in the vicinity of $u_s=0$ (equivalently, it has a
low-temperature series expansion with finite radius of convergence), so one
knows that all of the $4s^2$ curves which emanate from $u_s=0$ must move away
from the origin.  Evidently, the inner branches of the $n=0$ curve and its
complex conjugate ($n=4s^2-1$) curve move to the right and join at the point
$(u_s)_c = e^{-K_c/s^2}$ which constitutes the critical point between the FM
and PM phases; given the $u_s \to 1/u_s$ symmetry, this means that the outer
branches no longer run separately to complex infinity but join each other at
the critical point $1/(u_s)_c$ separating the PM and AFM phases of the 2D
model. This leaves $[s^2]-1$ curves in the first (NE) quadrant, which form
finite arcs; the inner and outer branches of these arcs protrude into and
terminate in the FM and AFM phases respectively (these branches and phases
are interchanged by the $u_s \to 1/u_s$ mapping). The situation in the complex
conjugate fourth (SE) quadrant is the same, by the $u_s \to u_s^*$ symmetry of
the phase diagram.  In the second (NW) quadrant, the $[s^2]$ curves in 1D
correspond to the $[s^2]$ arcs in 2D, which again have inner and outer branches
terminating in the FM and AFM phases.  The symmetry under complex conjugation
implies the same for the third (SW) quadrant.

    In summary, we have found an intriguing connection between the
complex-temperature phase diagrams, as calculated exactly for 1D and inferred
from partition function zeros for 2D, of the Ising model with higher spin.  Our
derivations here explain the basis for the conjecture that we made in
Ref. \cite{hs} on the total number of arc endpoints in the FM phase and hence
complex-temperature divergences in the magnetization for the square--lattice
Ising model with arbitrary spin.

This research was supported in part by the NSF grant PHY-93-09888.

\vfill
\eject

\begin{center}

{\bf Figure Captions}

\end{center}

 Fig. 1. Complex-temperature phase diagrams in the $u_s$ plane for the 1D
Ising model with (a) $s=1$ (with $u \equiv u_1$) and (b) $s=3/2$ (with $u
\equiv u_{3/2}$).  The phase boundaries are shown as the dark curves (including
the line along the negative $u_s$ axis for $s=3/2$).  The unit circle is drawn
in lightly for reference.

\vfill
\eject

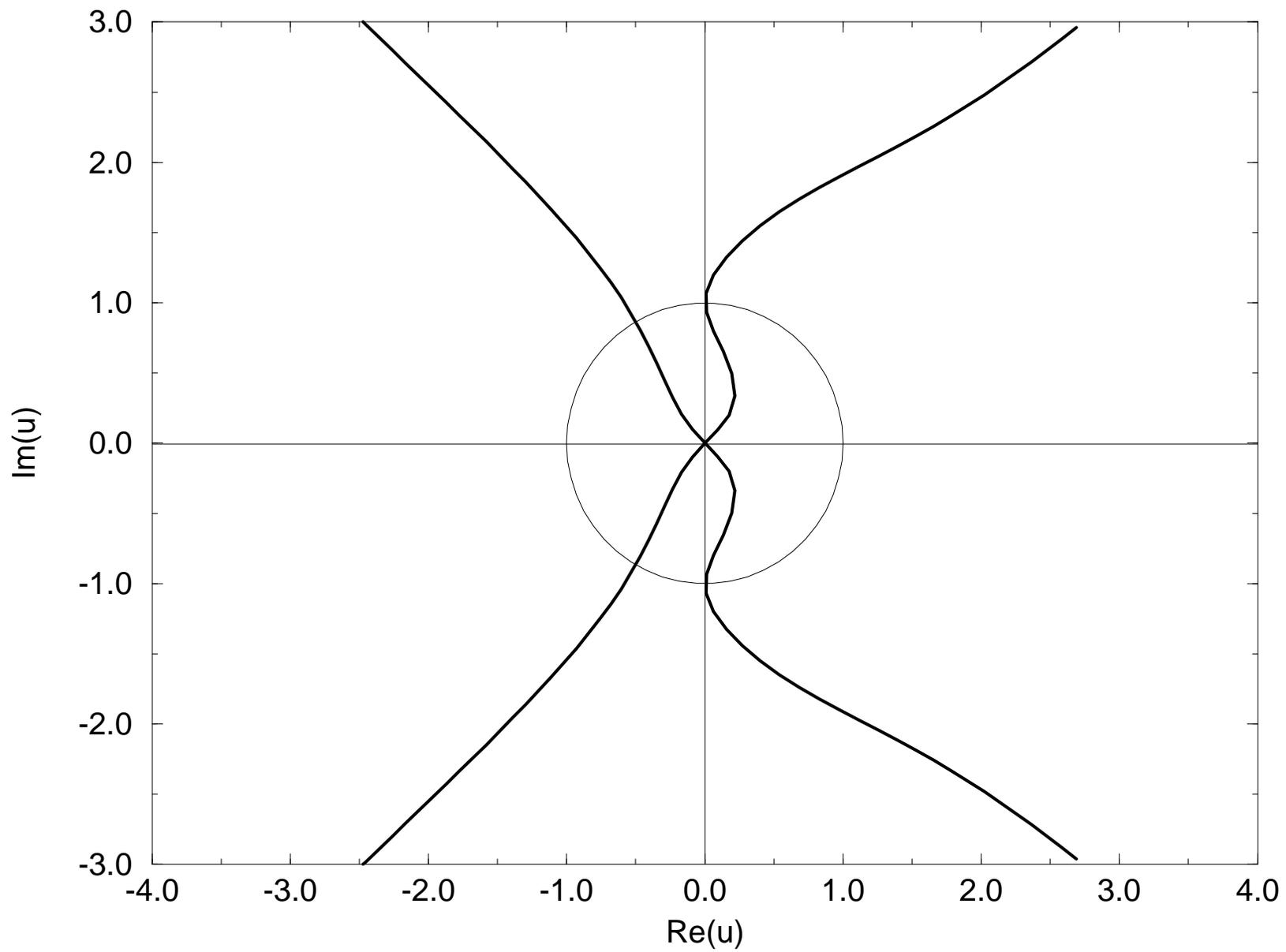
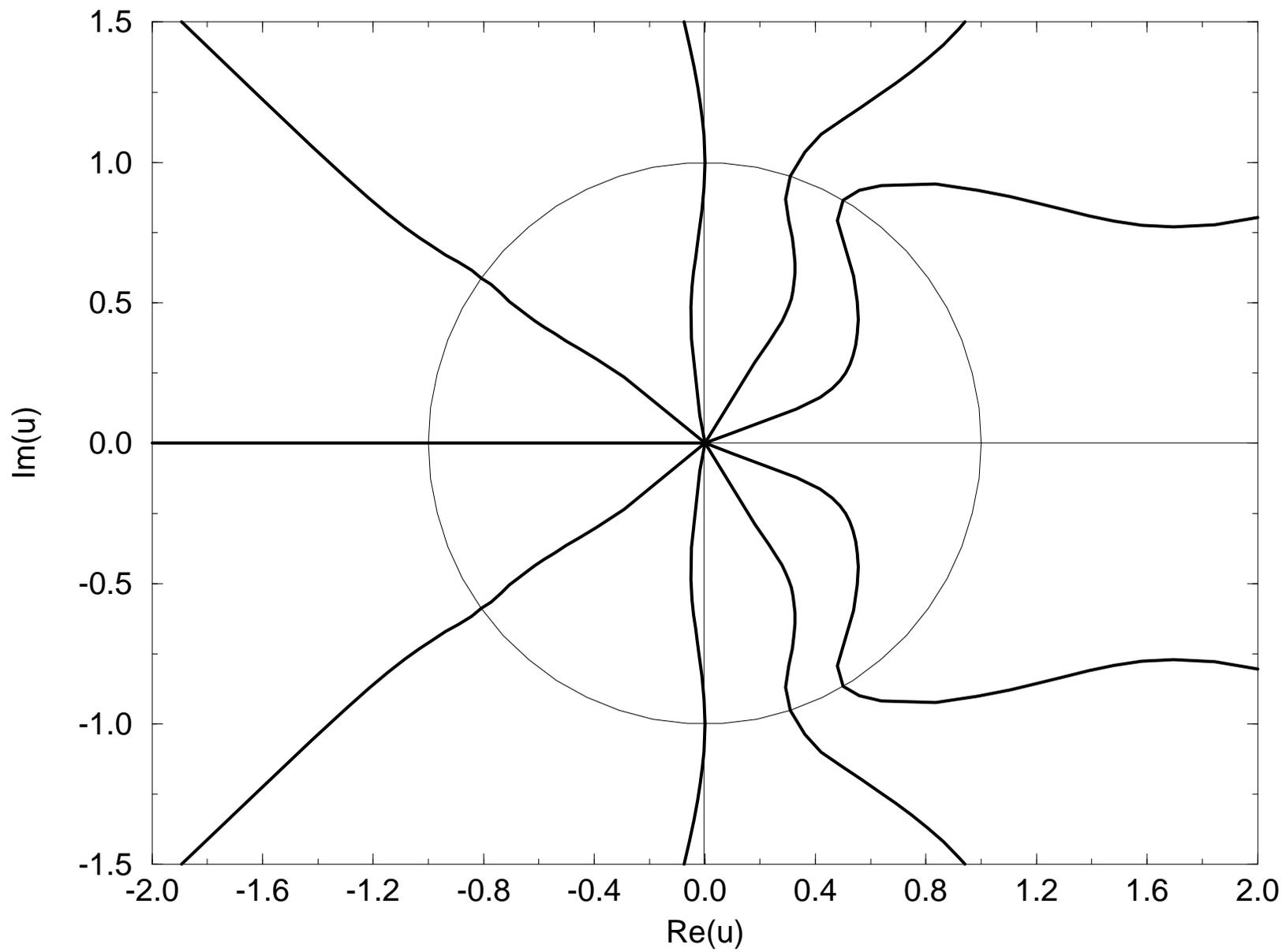
\end{document}